\documentclass[a4paper,12pt]{article}
\usepackage{latexsym}
\usepackage{amsfonts,amsbsy}
\newcommand{\be}{\begin{equation}}
\newcommand{\ee}{\end{equation}}
\newcommand{\bea}{\begin{eqnarray}}
\newcommand{\eea}{\end{eqnarray}} 

\usepackage{amsmath,amsfonts,amssymb,amscd,graphicx}
\catcode`\@=11
\catcode`\@=12

\newfont{\gotico}{eufm10 scaled\magstephalf}
\newfont{\qvd}{msam10 scaled\magstephalf}

\def\de#1/de#2{\frac{\partial {#1}}{\partial {#2}}}
\def\De#1/de#2{\dfrac{\partial {#1}}{\partial {#2}}}

\def\beqa{\begin{eqnarray}}
\def\eeqa{\end{eqnarray}}
\def\beq{\begin{equation}}
\def\eeq{\end{equation}}

\renewcommand{\epsilon}{\varepsilon}

\def\bet{\begin{tabular}}
\def\eet{\end{tabular}}
\def\bef{\begin{figure}}
\def\eef{\end{figure}}
\def\beqa{\begin{eqnarray}}
\def\eeqa{\end{eqnarray}}
\def\beq{\begin{equation}}
\def\eeq{\end{equation}}

\renewcommand{\epsilon}{\varepsilon}

\def\beqa{\begin{eqnarray}}
\def\eeqa{\end{eqnarray}}
\def\beq{\begin{equation}}
\def\eeq{\end{equation}}

\def\be{{\beta}}

\def\prl{{ Phys. Rev. Lett.}\ }

\begin{document}
 \title{Torsion, Scalar Field and $f(\mathcal{R})$ Gravity}
 \author{Prasanta Mahato 
        \thanks{email:pmahato@dataone.in} \\Narasinha Dutt College\\
         Howrah, West Bengal, India 711 101}

\date{}
\maketitle
\setcounter{footnote}{0}
 \begin{abstract}
 The role of  torsion and a scalar field $\phi$
   in gravitation in the background of a particular class of the Riemann-Cartan geometry is considered here.
 Some times ago,  a Lagrangian density with Lagrange multipliers has been proposed by the author which has been obtained  by picking some particular terms from the $SO(4,1)$ Pontryagin  density,  where the scalar field $\phi$ causes  the de Sitter connection to have the proper dimension of a gauge field. Here it has been shown that   the divergence of the axial torsion gives the Newton's constant and the scalar field  becomes a function of the Ricci scalar $\mathcal{R}$. The starting Lagrangian then reduces to a Lagrangian representing the metric $f(\mathcal{R})$  gravity theory.

\vspace{2mm}

\vspace{2mm}
   \noindent \textbf{\uppercase{key words :}}   Nieh-Yan Density, Scalar Field, Dark Matter, $f(\mathcal{R})$ Gravity\\
   \vspace{2mm}
    \vspace{2mm}
\noindent \textbf{\uppercase{pacs numbers :}} 04.20.Cv, 04.20.Fy, 98.80-k
 \end{abstract}
\section{Introduction}

Modern astrophysical and cosmological models are faced with two of the most fundamental theoretical problems of XXI century which are, namely,  the dark energy and the
dark matter problems. 
The galactic rotation curves of spiral galaxies~\cite{Tre87,Ste96,Sal01,Kle07}, probably, indicates the possible failure of Newtonian gravity and of
the general theory of relativity on galactic and intergalactic scales. In
these galaxies, neutral hydrogen clouds,  at large distances from
the center and  much beyond the extent of the luminous matter, found to be moving in circular orbits with nearly constant tangential velocity $v_{%
\mathrm{tg}}$. This yields an expression for the galactic mass profile of the
form $M(r)=rv_{\mathrm{tg}}^{2}/G$, with the total mass $M(r)$ increasing linearly with $r
$, even at large distances\cite{Tre87,Ste96,Sal01}. This bizarre behavior of the rotation curves makes the  dark matter postulate to be meaningful.

There are many possible candidates for dark matter\cite{Zac07}. However, no non-gravitational evidence for
the existence of dark matter has been reported so far.   In this context, several theoretical models,
based on a modification of Newton's law or of general relativity, at galactic scale, have been
proposed so far to explain the behavior of the galactic rotation curves~\cite
{Mil83,San84,Sok96,Man97,Rob04,Mof06,Mof06a,Har04,Che06,Har07,Che07,Bek04}.

Dark energy models mainly rely on the implicit assumption that
Einstein's GR is the correct theory of gravity indeed.
Nevertheless, its validity at the larger astrophysical and
cosmological scales has never been tested \cite{wil93}, and it is
therefore conceivable that both cosmic speed up and dark matter
represent signals of a breakdown in our understanding of
gravitation law so that one should consider the possibility that
the Einstein-Hilbert Lagrangian, linear in the Ricci scalar
$R$, should be generalized.

Following this line of thinking, the choice of a generic function
$f(\mathcal{R})$ can be derived by matching the data and by the 
requirement that no exotic ingredient have to be added. This is
the underlying philosophy of what is referred to as $f(\mathcal{R})$ gravity
\cite{Odi07,Cap02,Odi03,Odi03a,Tro03,Tur04,Tur05,Fla03,Fra04,Tro05a,Koi06,Buc70,Ott83}. It has been suggested that
these modified gravity models account for the late time
acceleration of the universe~\cite{Tur04}, thus
challenging the need for dark energy. Though, severe weak field constraints in the solar system range
may  rule out many of the models proposed so
far~\cite{Chi03,Kam06,Eri07,Olm07} but there exists some definite viable models~\cite
{Saw07,Odi03a,Far06,Mao07,Hu07,Tsu07,Bar07}.
In addition to satisfy the solar system
constraints, the viable models should simultaneously account for the four
distinct cosmological phases, namely, inflation, the radiation-dominated and
matter-dominated epochs, and the late-time accelerated expansion~\cite
{Odi06,Odi06a,Tsu07b,Odi07a}, and be consistent with cosmological structure formation
observations~\cite{Kur06,Tro07,Bar07,Chu06,Chu07}. The issue of stability~\cite
{Far06,Tro05,Dun06,Dun07,Bar07} also plays an important role in the viability of
cosmological solutions~\cite{Hu07,Tsu07c,Odi07,Sok07,Zer05,Far05a,Far05b,Far07,Zer07,Lob07a,Tro07a,Dun07a}.
It is interesting to note that, recently,  it has been found that same $f(\mathcal{R})$ gravity models satisfying
cosmological and local gravity constraints are practically indistinguishable
from the $\Lambda$CDM model, at least at the background level~\cite
{Tsu07}.

In the context of galactic dynamics, a version of $f(\mathcal{R})$ gravity
models admitting a modified Schwarzschild-de Sitter metric has been
analyzed in~\cite {Sob07}. In the weak field limit one obtains a
small logarithmic correction to the Newtonian potential, and a
test star moving in such a spacetime acquires a constant
asymptotic speed at large distances. It is interesting to note
that the model has similar properties with MOND~\cite{Mil83,San84,Bek04}. A
model based on a generalized action with $f(R) = R + R(R/R_0 +
2/\alpha)^{-1}\ln(R/R_c)$, where $\alpha $, $R_0$ and $R_c$ are
constants, was proposed in~\cite{Rah07}. In particular, this
model can describe the Pioneer anomaly\cite{And05} and the flat rotation
curves of the spiral galaxies. In a cosmological context, the
vacuum solution also results in a late time acceleration for the
universe.

  It is a remarkable result of differential geometry that
 certain global features of a manifold are determined by some local invariant densities.
These topological invariants have  an important property in common -  they are total divergences
and in any local theory  these invariants,  when treated as Lagrangian densities, contribute
 nothing to the Euler-Lagrange equations.  Hence in a local theory only few parts, not the
  whole part, of these invariants can be kept in a Lagrangian density. Recently, in
this direction, a gravitational Lagrangian has been
proposed\cite{Mah02a}, where a
 Lorentz invariant part of the de Sitter Pontryagin density has been treated as
 the Einstein-Hilbert Lagrangian.  By this way the role of torsion in the underlying manifold
has become multiplicative
   rather than additive one and  the  Lagrangian  looks like
    $\mathbf{torsion \otimes curvature}$.  In other words - the additive torsion is decoupled from the
   theory but not the multiplicative one. This indicates that torsion is uniformly nonzero
   everywhere. In the geometrical sense, this implies that
   micro local space-time is such that at every point there is a
   direction vector (vortex line) attached to it. This effectively
   corresponds to the non commutative geometry having the manifold
   $M_{4}\times Z_{2}$, where the discrete space $Z_{2}$ is just
   not the two point space\cite{Con94} but appears as an attached direction vector. This has direct relevance in the quantization of a fermion where the discrete space appears as the internal space of a particle\cite{Gho00}. Considering torsion and torsion-less
  connection as independent fields\cite{Mah04}, it has been found that $\kappa$  of Einstein-Hilbert Lagrangian, appears as an integration constant in such a way that it has been found to be linked with the topological Nieh-Yan density of $U_{4}$ space.
  If we consider axial vector torsion together with a scalar field $\phi$ connected to a local  scale factor\cite{Mah05}, then the Euler-Lagrange equations not only give the constancy of the gravitational constant but they also link, in laboratory scale, the mass of the scalar field with the Nieh-Yan density and, in cosmic scale of FRW-cosmology, they predict only three kinds of the phenomenological energy density representing mass, radiation and cosmological constant. In a recent paper\cite{Mah07}, it has been shown that this scalar field may also be interpreted to be linked with the dark matter and dark radiation.

Up to some time ago, torsion  did not seem to produce models with
observable effects since  phenomena implying  spin and gravity
were considered to be significant only in the very early Universe.
After, it has been proved  that spin  is not the only source of
torsion\cite{Sto01}. This means that a wide class of torsion models could be
investigated  independently of spin as their source.

In this paper, following the same philosophy, we want to show
that, starting from a Lagrangian of the type in Refs.\cite{Mah05,Mah07}, how a generic $f(\mathcal{R})$ gravity theory emerges. Such that, the curvature, torsion and the
scalar field may give rise to an effective $f(\mathcal{R})$ gravity theory which is capable, in principle, to address the problem
of the Dark Side of the Universe in a very general geometric
scheme. 

The layout of the paper is the following. In Sec.II we briefly describe the geometry and the starting Larangian  in the background of a particular class of Riemann-Cartan geometry of the space-time manifold. In Sec.III, we
derive the generic $f(\mathcal{R})$ gravity Lagrangian together with some  particular cases. 
Sec.IV is devoted to some discussion.


 \section{Axial Vector Torsion and Scalar Field}

Cartan's structural equations for a Riemann-Cartan space-time $U_{4}$ are given by
\cite{Car22,Car24}
 \begin{eqnarray}T^{a}&=& de^{a}+\omega^{a}{}_{b}\wedge e^{b}\label{eqn:ab}\\
 R^{a}{}_{b}&=&d\omega^{a}{}_{b}+\omega^{a}{}_{c}\wedge \omega^{c}{}_{b},\label{eqn:ac}
\end{eqnarray}
here $\omega^{a}{}_{b}$ and e$^{a}$ represent the spin connection
and the local frame respectively.

In $U_{4}$ there exists  two invariant closed four forms. One is the well
known Pontryagin\cite{Che74,Che71} density \textit{P} and the
other is the less known Nieh-Yan\cite{Nie82} density \textit{N}
given by
\begin{eqnarray} \textit{P}&=& R^{ab}\wedge R_{ab}\label{eqn:ad}\\  \mbox{and} \hspace{2 mm}
 \textit{N}&=& d(e_{a}\wedge T^{a})\nonumber\\
&=&T^{a}\wedge T_{a}- R_{ab}\wedge e^{a}\wedge
e^{b}.\label{eqn:af}\end{eqnarray}

Here we consider a particular class of the Riemann-Cartan geometry where only the axial vector part of the torsion is nontrivial.
Then, from  (\ref{eqn:af}),  one naturally gets the Nieh-Yan
density 
\begin{eqnarray} N&=&-R_{ab}\wedge e^{a}\wedge e^{b}=-{}^* N\eta\hspace{2 mm},
\label{eqn:xaa}\\
 \mbox{where} \hspace{2 mm}\eta&:=&\frac{1}{4!}\epsilon_{abcd}e^{a}\wedge e^b\wedge
e^c\wedge e^d\end{eqnarray}is the invariant volume element.  It follows that    ${}^*N$,
the Hodge dual of $N$, is a scalar density of dimension $(length)^{-2}$.

We can combine the spin connection and the vierbeins multiplied by a scalar field together in a connection for $SO(4,1)$, in the tangent space, in the
form
\begin{eqnarray}W^{AB}&=&\left
[\begin{array}{cc}\omega^{ab}&\varphi e^{a}\\- \varphi e^{b}&0\end{array}\right],\label{eqn:aab}
\end{eqnarray}
where $a,b = 1,2,..4$; $A,B = 1,2,..5$ and $\varphi$ is a variable
parameter of dimension $(length)^{-1}$ and Weyl weight $(-1)$, such
that, $\varphi e^a$ has the correct  dimension and conformal weight
of the de Sitter boost part of the $SO(4,1)$ gauge connection. In
some earlier works\cite{Cha97,Mah02a,Mah04}  $\varphi$ has been
treated as an inverse length constant. In another earlier work\cite{Mah05} $\varphi$ has been associated, either in laboratory
scale or in cosmic sale, with a local energy scale. In laboratory
scale its coupling with torsion gives the mass term of the scalar
field and in cosmic scale it exactly produces the phenomenological
energy densities of the FRW universe. In a recent paper\cite{Mah07} the scalar field $\phi$ is associated with the dimension of a spinor field $\Psi$ and is found that $\phi$ does not interect with $\Psi$ and thus the scalar field may  be representing the dark matter and(or) dark radiation.  In this line of approach, the gravitational Lagrangian with only a scalar field $\varphi$, may be proposed  to be
\begin{eqnarray}\mbox{$\mathcal{L}$}_{G}&=& \frac{1}{6}({}^*N \mbox{$\mathcal{R}$}\eta  +\beta \varphi^2 N)+{}^*(b_a\wedge
\bar{\nabla} e^{a})(b_a\wedge
\bar{\nabla} e^{a}) \nonumber\\&{}&- \frac{1}{2}w(\phi)d\varphi\wedge{}^*d\varphi +\widetilde{h}(\varphi)
\eta,\label{eqn:abcd1}\end{eqnarray} where
  *   is Hodge duality operator,    $\mathcal {R}$$\eta=\frac{1}{2}\bar{R}^{ab}\wedge\eta_{ab}$, $\bar{R}^b{}_a=d\bar{\omega}^b{}_a+\bar{\omega}^b{}_c\wedge \bar{\omega}^c{}_a$, $\bar{\omega}^{a}{}_{b}=\omega^{a}{}_{b}-T^{a}{}_{b}$, $ T^a=e^{a\mu}T_{\mu\nu\alpha}dx^\nu\wedge dx^\alpha$, $T^{ab}=e^{a\mu}e^{b\nu}T_{\mu\nu\alpha}  dx^\alpha$, $T=\frac{1}{3!}T_{\mu\nu\alpha}dx^\mu\wedge dx^\nu\wedge dx^\alpha$, $N=6dT$, $\eta_a=\frac{1}{3!}\epsilon_{abcd}e^b\wedge e^c\wedge e^d$ and $\eta_{ab}={}^*(e_a\wedge e_b)$.  Here $\beta$ is a dimensionless coupling constant, $\bar{\nabla}$
represents covariant differentiation with respect to the connection one form $\bar{\omega}^{ab}$, $b_{  a}$ is a two form with
one internal index and of dimension $(length)^{-1}$ and $w(\varphi)$,  $\widetilde{h}(\varphi)$ are unknown functions of $\varphi$ whose
forms are to be determined subject to the geometric structure of the manifold.
  The geometrical implication of the first term, i.e. the $\mathbf{torsion \otimes curvature}$\footnote{An important advantage of this part of the Lagrangian is that - it is a
   quadratic one with respect to the field derivatives and this
   could be valuable in relation to the quantization program of gravity like other gauge theories of
   QFT.} term, in the Lagrangian $\mathcal{L}\mbox{$_{G}$}$   has already been  discussed in the beginning.

  The Lagrangian $\mathcal{L}\mbox{$_{G}$}$
     is  only Lorentz invariant  under rotation in the tangent space where  de Sitter
     boosts are not permitted. As a consequence $T$ can be treated independent of $e^a$
     and $\bar{\omega}^{ab}$.
Here we note that, though torsion one form $T^{ab}=\omega^{ab}-\bar{\omega}^{ab}$ is a part of
the $SO(3,1)$ connection, it
 does not transform like a connection  form under $SO(3,1)$   rotation in
 the tangent space  and thus it imparts no constraint on the gauge degree of freedom of the
   Lagrangian.
 
 Following Refs. \cite{Odi07, Odi06b, Sok94} we define a new scalar field $\phi$ as
\begin{eqnarray}
	\phi=\int d\varphi\sqrt{|w(\varphi)|}\label{eqn:abcd11}\hspace{1mm},
\end{eqnarray}then the lagrangian (\ref{eqn:abcd1}) becomes 
\begin{eqnarray}
	\mbox{$\mathcal{L}$}_{G}&=& \frac{1}{6}\{{}^*N \mbox{$\mathcal{R}$}\eta  +\beta u(\phi) N\}+{}^*(b_a\wedge
\bar{\nabla} e^{a})(b_a\wedge
\bar{\nabla} e^{a}) \nonumber\\&{}&\mp \frac{1}{2}d\phi\wedge{}^*d\phi +h(\phi)
\eta,\label{eqn:abcd12}
\end{eqnarray}here the sign in front of the kinetic term depends on the sign of $w(\varphi)$ and by Eqn. (\ref{eqn:abcd11}) we can express $\phi$ as a function of $\varphi$, i.e.,  $\phi\equiv\phi(\varphi)$,  such that we can also define two functions $u(\phi)$ and $h(\phi)$ by
\begin{eqnarray}
	u(\phi)= u(\phi(\varphi))\equiv\varphi^2\hspace{4mm}\mbox{and}\hspace{4mm}h(\phi)= h(\phi(\varphi))\equiv\widetilde{h}(\varphi)\label{eqn:abcd13}
\end{eqnarray}

 \section{Scalar Field and $f(\mathcal{R})$ Gravity}

In  appendix A,  by varying  the independent fields except the frame field $e^a$ in the Lagrangian $\mathcal{L}\mbox{$_{G}$}$,  we obtain the Euler-Lagrange equations and then after some simplification we get the following results
\begin{flushright}$\begin{array}{lcr}
&\bar{\nabla} e_{a}=0,\hspace{38mm}&(\ref{eqn:abc9}^\prime)\nonumber\\&
{}^*N=\frac{6}{\kappa},\hspace{38mm}&(\ref{eqn:abc17}^\prime)\nonumber
\end{array}$\end{flushright}
 i.e. $\bar{\nabla}$ is torsion free and $\kappa$ is an integration constant having  dimension of $(length)^{2}$\footnote{In (\ref{eqn:abcd1}), $\bar{\nabla}$ represents a $SO(3,1)$ covariant derivative, it is only on-shell torsion-free through the field equation (\ref{eqn:abc9}$^\prime$). This amounts to the emergence of the gravitational constant   $\kappa$ to be  only an on-shell  constant and this justifies the need for the introduction of the Lagrangian multiplier $b_a$ which appears twice in the Lagrangian density (\ref{eqn:abcd1}) such that $\bar{\omega}^a{}_b$ and $e^a$  become independent fields.}.

 From equation (\ref{eqn:abc14}) we can write 
\begin{eqnarray}
	\mbox{$\mathcal{R}$}-\beta u(\phi)=\lambda, \label{eqn:abcd17}
\end{eqnarray}the integration constant $\lambda$ has been associated with dark energy in references \cite{Mah05, Mah07}.

 Using equation (\ref{eqn:abc17}$^\prime$), equation (\ref{eqn:abc140}) can be written as
\begin{eqnarray}
\pm d{}^*d\phi&=& \{\frac{1}{\kappa}\beta u^\prime{(\phi)}-h^\prime(\phi)\}
\eta\nonumber\\\mbox{or,}\hspace{4mm}d\phi\wedge{}^*d\phi&=&d(\phi{}^*d\phi)\mp(\frac{1}{\kappa}\beta u^\prime-h^\prime)\phi\eta\label{eqn:abc1400}
 \end{eqnarray}
 
 Now the Lagrangian (\ref{eqn:abcd1}) can be written as
 \begin{eqnarray}\mbox{$\mathcal{L}$}_{G}&=& \frac{1}{\kappa}( \mbox{$\mathcal{R}$}-\beta u )\eta\mp \frac{1}{2}\{d(\phi{}^*d\phi)\mp(\frac{1}{\kappa}\beta u^\prime-h^\prime)\phi\eta\}+h\eta\nonumber\\&=&\frac{1}{\kappa}\{ \mbox{$\mathcal{R}$}-\beta u +\frac{1}{2}(\beta u^\prime-\kappa h^\prime)\phi+\kappa h\}\eta\mp\frac{1}{2}d(\phi{}^*d\phi)\nonumber\\&=&\frac{1}{\kappa}\{ \mbox{$\mathcal{R}$}-V(\phi)\}\eta\mp\frac{1}{2}d(\phi{}^*d\phi),\label{eqn:abcde1}\\\mbox{where,} 
\hspace{4mm}V(\phi)&=&(\beta u-\kappa h)-\frac{1}{2}(\beta u^\prime-\kappa h^\prime)\phi\label{eqn:abcde2}
\end{eqnarray}

From equation (\ref{eqn:abcd17}) we can express $\phi$ as a function of $\mathcal{R}$, i.e., $\phi\equiv\phi(\mathcal{R})$ and then the  Lagrangian (\ref{eqn:abcde1}) takes the following form
\begin{eqnarray}
	\mbox{$\mathcal{L}$}_{G}&=& \frac{1}{\kappa}f(\mathcal{R})\eta+ (\mbox{surface term})\label{eqn:abcde3}\\\mbox{where,}\hspace{4mm}f(\mathcal{R})&=&\mathcal{R}-V(\phi(\mathcal{R}))\label{eqn:abcde4}
\end{eqnarray}
Hence the Lagrangian (\ref{eqn:abcd1}) reduces to a  $f(\mathcal{R})$ gravity Lagrangian and due to equation ($\ref{eqn:abc9}^\prime$), this Lagrangian gives the dynamics of a metric $f(\mathcal{R})$ gravity theory.

Varying the action corresponding to the Lagrangian (\ref{eqn:abcde3}) with respect to the metric $g_{\mu \nu }$ yields the
following field equations
 \begin{align}
f^\prime(\mathcal{R})R_{\mu \nu }-\frac{1}{2}f(\mathcal{R})g_{\mu \nu }-\left(\nabla _{\mu }\nabla
_{\nu }-g_{\mu \nu }\square \right)f^\prime(\mathcal{R})=0,  \label{field}
\end{align}
where we have denoted $f^\prime(\mathcal{R})=d f(\mathcal{R})/d \mathcal{R}$. This equation can also be written as
\begin{eqnarray}
	G_{\mu\nu}= \frac{1}{2}g_{\mu\nu}(\frac{f}{f^\prime}-\mathcal{R})+\frac{1}{f^\prime}(\nabla_\mu\nabla_\nu-\nabla_\mu\nabla^\mu)f^\prime\equiv\kappa T_{\mu\nu}\hspace{2mm}\mbox{(say)} \label{tensor01}
\end{eqnarray} 

Using the results of Ref. \cite{Mah05} and the definition (\ref{eqn:abcd13}) of $u(\phi)$ and $h(\phi)$ as functions of $\phi$ we see that, in the FRW background, where the metric is given by 
\begin{eqnarray}
	g_{00}=-1 ,\hspace{2mm}g_{ij}= \delta_{ij}a^2(t)  \hspace{2mm}\mbox{where}\hspace{2mm}
i,j=1,2,3;\label{eqn:aay}
\end{eqnarray}the total energy density can be wreitten as 
\begin{eqnarray}
	\rho=\rho_{{}_M}+\rho_{{}_R}+\rho_{{}_{VAC.}}
	\end{eqnarray}
	where $h=-\gamma  u^\frac{4}{3}+\frac{\lambda}{2\kappa}$ and
	\begin{enumerate}
\item the pressure-less mass density $\rho_{{}_M}=\frac{\beta}{\kappa} u\propto
a^{-3}$,
\item the radiation density $\rho_{{}_R}=\frac{3}{2}\gamma u^{\frac{4}{3}}\propto
a^{-4}$ where pressure $p_{{}_R}=\frac{1}{3}\rho_{{}_R}$ and
\item the constant vacuum energy density $\rho_{{}_{VAC.}}=\frac{\lambda}{4\kappa}$
where pressure $ p_{{}_{VAC.}}=-\rho_{{}_{VAC.}}$.
\end{enumerate}

 Now, since the metric $f(\mathcal{R})$ gravity Lagrangian (\ref{eqn:abcde3}) is obtained by eliminating the non-metrical field variables of the Lagrangia (\ref{eqn:abcd12}), it will also give us the same standard FRW cosmology with  only three specific  kinds of energy densities when the background FRW metric (\ref{eqn:aay}) is used\cite{Mah05}. That is, we can write  $T_{00}=\rho_{{}_M}+\rho_{{}_R}+\rho_{{}_{VAC.}}$, where $T_{00}$ is defined in (\ref{tensor01}). It is surprising that these are the only three kinds of phenomenological cosmic energy density that we observe and consider to be interested in. But theoretically, in standard FRW cosmology, other forms of energy density are not ruled out\cite{Mah05}. And therefore to consider other forms of cosmic energy density in the early universe or in the galactic scale, we have to adopt a non-FRW geometry where we may have to forgo the isotropy and (or) the homogeneity of the universe. In particular, here in the following, let us consider a non-FRW geometry in galactic scale. 

B\"ohmer et. al.\cite{Lob07b}, to address dark matter problem, have considered the galactic dynamics by restricting the study to the static and spherically symmetric metric given by 
\begin{eqnarray}
	ds^2=-e^{\nu(r)}dt^2+e^{\lambda(r)}dr^2+r^2d\Omega^2.\label{mond01}
\end{eqnarray}where $d\Omega^2=d\theta^2+\sin^2\theta d\phi^2$. They have found that in the flat velocity curves region the metric coefficients are given by 
\begin{eqnarray}
	\nu=2v^2_{tg}\ln{(\frac{r}{r_0})}\hspace{2mm}\mbox{and}\hspace{2mm}e^\lambda \approx1+2v^2_{tg},
\end{eqnarray}here $v_{tg}$ is the constant tangential velocity of the stars and gas clouds in circular orbits in the outskirts of spiral galaxies. In the limit of large $r$, the Newtonian potential is given by $\Phi_N(r)\approx v^2_{tg}\ln{(\frac{r}{r_0})}$, reflecting a logarithmic dependence on the radial distance $r$. Therefore having a well-defined Newtonian limit the metric (\ref{mond01}) can be used to describe the geometry of the space and time in the dark matter dominated regions. They have shown that, in the $f(\mathcal{R})$ gravity models, the rotational galactic curves can be naturally explained without introducing any additional hypothesis, by taking
\begin{eqnarray}
	f(\mathcal{R})=f_0\mathcal{R}^{1+v^2_{tg}}\label{mond02}
\end{eqnarray}where $f_0$ is a positive constant.
 
 We see that if we take $\beta u=(2\alpha f_0\ln{\frac{\phi_0}{\phi}})^{-\frac{1}{\alpha}}$ and $h=0$, where $\alpha=v^2_{tg}$ and $\phi_0$ a positive constant, then $f(\mathcal{R})$ of Eqn. (\ref{eqn:abcde4}) coincides with that of Eqn. (\ref{mond02}).
 
\section{Discussion} In this article, we have seen that if we introduce a scalar field $\phi$ to cause  the de Sitter connection to have the proper dimension of a gauge field and  then vary the $SO(3,1)$ spin connection as an entity independent of the tetrads, we get the Newton's constant as inversely proportional to the topological Nieh-Yan density. Then  Euler-Lagrange equations of the axial torsion $T$ and the scalar field $\phi$ reduce the Lagrangian to that of the metric $f(\mathcal{R})$ gravity.

 In this present analysis it is significant that if we consider our universe to have the isotropy and the homogeneity of a FRW universe then only three kinds of energy densities are possible. The matter energy density $\propto a^{-3}$, the radiation energy density $\propto a^{-4}$ and the vacuum energy density $\propto a^0$ are the only three kinds of such energy densities where $a$ is the cosmic scale factor. It is surprising that these are the only three kinds of phenomenological cosmic energy density that we observe and consider to be interested in. But theoretically, in standard FRW cosmology, other forms of energy density are not ruled out\cite{Mah05}. And therefore to consider other forms of cosmic energy density we have to adopt a non-FRW geometry. 
 
 In galactic scale B\"ohmer et. al.\cite{Lob07b} adopted a non-FRW metric and obtained flat galacic rotation curves for spiral galaxies. The corresponding $f(\mathcal{R})$ gravity Lagrangian was found to be proportional to $\mathcal{R}^{1+v_{tg}^2}$, where $v_{tg}$ was the constant tangential velocity in the flat rotation curves region around spiral galaxies. 
 
In our present formalism, the starting Lagrangian (\ref{eqn:abcd12}) reduces to a generic $f(\mathcal{R})$ gravity Lagrangian (\ref{eqn:abcde3}) which, for FRW metric,  gives standard FRW cosmology. But for non-FRW metric, in particular of Ref.\cite{Lob07b}, with some particular choice of the functions $u$ and $h$ one gets $f(\mathcal{R})=f_0\mathcal{R}^{1+v^2_{tg}}$. With this choice of $u$ and $h$ no dark matter is required to explain flat galactic  rotation curves.
 
 In conclusion, we can say that, with certain choice of the metric and the functions $u$ and $h$, which are in conformity with themselves, one may get some specific forms  of the function $f(\mathcal{R})$. This may be used to explain some of the anomalous features of the universe. Further investigation may be done in a future article.
 \section*{Acknowledgment}
           I wish to thank Prof. Pratul Bandyopadhyay, Indian Statistical Institute,  Kolkata,  for his valuable remarks and fruitful suggestions on this
           problem.

\renewcommand{\theequation}{\mbox{A}\arabic{equation}}
\setcounter{equation}{0}

\begin{appendix}
\section*{Appendix A}
 Following reference \cite{Heh95}, we independently vary
     $dT$,  $\bar{R}^{ab}$, $\phi$, $d\phi$ and
    $b^{a}$  and find

\begin{eqnarray}
    \delta  \mathcal{L}\mbox{$_{G}$}&=&\delta dT \frac{\partial  \mathcal{L}\mbox{$_{G}$}}{\partial
    dT}+\delta \bar{R}^{ab}\wedge \frac{\partial  \mathcal{L}
\mbox{$_{G}$}}{\partial
    \bar{R}^{ab}}+\delta\phi\frac{\partial  \mathcal{L}\mbox{$_{G}$}}{\partial \phi}\nonumber\\&{}&+\delta
    d\phi\wedge\frac{\partial  \mathcal{L}\mbox{$_{G}$}}{\partial d\phi}+\delta b^a\wedge
    \frac{\partial  \mathcal{L}\mbox{$_{G}$}}{\partial b^a}\nonumber \\&=&\delta T\wedge d \frac{\partial  \mathcal{L}
    \mbox{$_{G}$}}{\partial dT}+\delta \bar{\omega}^{ab}\wedge(\bar{\nabla} \frac{\partial
     \mathcal{L}\mbox{$_{G}$}}{\partial \bar{R}^{ab}}+ \frac{\partial
    \mathcal{L}\mbox{$_{G}$}}{\partial \bar{\nabla} e^a}\wedge e_b)\nonumber\\&{}&+\delta \phi(\frac{\partial
    \mathcal{L}\mbox{$_{G}$}}{\partial \phi}- d \frac{\partial  \mathcal{L}\mbox{$_{G}$}}
    {\partial d\phi})+\delta b^a\wedge \frac{\partial  \mathcal{L}\mbox{$_{G}$}}{\partial
    b^a}\nonumber \\&{}&
    +d(\delta T \frac{\partial  \mathcal{L}\mbox{$_{G}$}}{\partial dT}+\delta
    \bar{\omega}^{ab}\wedge \frac{\partial  \mathcal{L}\mbox{$_{G}$}}{\partial \bar{R}^{ab}}+
    \delta\phi\frac{\partial  \mathcal{L}\mbox{$_{G}$}}{\partial d\phi})\label{eqn:abc0}
\end{eqnarray}Using the form of the Lagrangian $\mathcal{L}\mbox{$_{G}$}$, given in (\ref{eqn:abcd1}),
we get
\begin{eqnarray}
    \frac{\partial  \mathcal{L}\mbox{$_{G}$}}{\partial (dT)}&=&-\mbox{$\mathcal{R}$}+\beta u(\phi)\label{eqn:abc4}\\\frac{\partial  \mathcal{L}\mbox{$_{G}$}}{\partial \bar{R}^{ab}}&=&\frac{1}{24}{}^*N\epsilon_{abcd}e^c\wedge e^d=\frac{1}{12}{}^*N\eta_{ab}\label{eqn:abc5}\\\frac{\partial  \mathcal{L}\mbox{$_{G}$}}{\partial \phi}&=&\frac{1}{6}\beta u^\prime(\phi) N+h^\prime(\phi)
\eta\label{eqn:abc5a}\\\frac{\partial  \mathcal{L}\mbox{$_{G}$}}{\partial d\phi}&=&\mp{}^*d\phi\label{eqn:abc5b}\\\frac{\partial  \mathcal{L}\mbox{$_{G}$}}{\partial b^a}&=&2{}^*(b_b\wedge
\bar{\nabla} e^{b})
\bar{\nabla} e_{a}\label{eqn:abc6}
\eea
here ${}^\prime$ represents derivative w.r.t. $\phi$.

From above, Euler-Lagrange equations for  $b_a$ gives us
\begin{eqnarray}
\bar{\nabla} e_{a}&=&0\label{eqn:abc9}
\end{eqnarray}i.e.   $\bar{\nabla}$ is torsion free.

Euler-Lagrange equations corresponding to the  extremum of $\mbox{$\mathcal{L}$}_{G}$ from the independent variations  of  $T$, $\phi$ and $\bar{\omega}^{ab}$, using (\ref{eqn:abc0}),  (\ref{eqn:abc4})  and (\ref{eqn:abc5}), give us
\begin{eqnarray}&{}&
d(\mbox{$\mathcal{R}$}-\beta u(\phi))=0\label{eqn:abc14}
\\&{}&\frac{1}{6}\beta u^\prime{(\phi)} N +h^\prime(\phi)
\eta\pm d{}^*d\phi=0\label{eqn:abc140}
\end{eqnarray}
\begin{eqnarray}
\bar{\nabla}({}^*N\eta_{ab})&=&0\label{eqn:abc15}
\end{eqnarray}
Using (\ref{eqn:abc9}) in (\ref{eqn:abc15}), we get
\begin{eqnarray}
    d{}^*N=0\label{eqn:abc16}
\end{eqnarray}
From  this equation  we can write
\begin{eqnarray}
{}^*N=\frac{6}{\kappa}\label{eqn:abc17}
\end{eqnarray}where $\kappa$ is an integration constant having $(length)^{2}$ dimension.

\end{appendix}
 \bibliographystyle{unsrt}              



 \end{document}